%% file: Paper_Main_IEEE_CP_Notice.tex
\documentclass[letterpaper, 10 pt, conference]{ieeeconf}  

\IEEEoverridecommandlockouts                              

\usepackage{amsfonts}
\let\labelindent\relax
\usepackage{enumitem}
\usepackage{import}
\usepackage{graphicx}

\def\FigPath{Figures/}

\input{MyCmd}

\graphicspath{{Figures/}}

\title{\LARGE \bf
Towards Real-Time Monitoring and Control of Water Networks
}

\author{Ahmed Elkhashap$^{1}$, Daniel Rüschen$^{2}$ and Dirk Abel$^{1}$
\thanks{$^{1}$Ahmed Elkhashap and Dirk Abel are with Institute of Automatic Control, RWTH Aachen University, Aachen, Germany {\tt\small a.elkhashap@irt.rwth-aachen.de}}%
\thanks{$^{2}$ Daniel Rüschen is with Viega GmbH \& Co. KG, Attendorn, Germany
        {\tt\small daniel.rueschen@viega.de}}%
}

\newcommand\copyrighttext{%
  \footnotesize \textcopyright 2022 IEEE. Personal use of this material is permitted.
  Permission from IEEE must be obtained for all other uses, in any current or future
  media, including reprinting/republishing this material for advertising or promotional
  purposes, creating new collective works, for resale or redistribution to servers or
  lists, or reuse of any copyrighted component of this work in other works.}
\newcommand\copyrightnotice{%
\begin{tikzpicture}[remember picture,overlay]
\node[anchor=south,yshift=10pt] at (current page.south) {\fbox{\parbox{\dimexpr\textwidth-\fboxsep-\fboxrule\relax}{\copyrighttext}}};
\end{tikzpicture}%
}
\begin{document}

\maketitle
\copyrightnotice
\thispagestyle{empty}
\pagestyle{empty}

\begin{abstract}

Water networks are used in numerous applications, all of which have the essential task of real-time monitoring and control of water states. A framework for the generation of efficient models of water networks suitable for real-time monitoring and control purposes is proposed. The proposed models preserve the distributed parameter character of the connected local elements. Hence, the spatial resolution of the property under consideration is recovered. The real-time feasibility of the network model is ensured by means of reduced order modeling of the models constituting components. A novel model order reduction procedure that preserves the model parametric dependency is introduced. The proposed concept is evaluated with the water temperature as the property under consideration. The formulated model is applied for the prediction of the water temperature within an experimental test bench of a 60 meter exemplary circulation network at the company VIEGA. A reduced order model (ROM) with a 50 mm spatial resolution, i.e. 1200 discretization points, is constructed and utilized as the identification model for a single path of the test bench. Afterwards, the ROM is evaluated in a Hardware in the Loop experiment for the prediction of the downstream temperature showing high prediction accuracy with mean relative error below 3.5 \%. The ROM single step computation time did not exceed 2 msec highlighting the real-time potential of the method. Moreover, full network model validation experiments featuring both diffusion and transport dominated parts were conducted. The network model is able to predict the temperature evolution, flow rate, and pressure accurately at the different paths of the network with mean relative errors below 4 \%, 2 \%, and 2 \%, respectively. \end{abstract}

\section{INTRODUCTION}
\begin{figure}[t!]
	\centering
	\vspace{-0.4cm}
	\includegraphics[width=\linewidth]{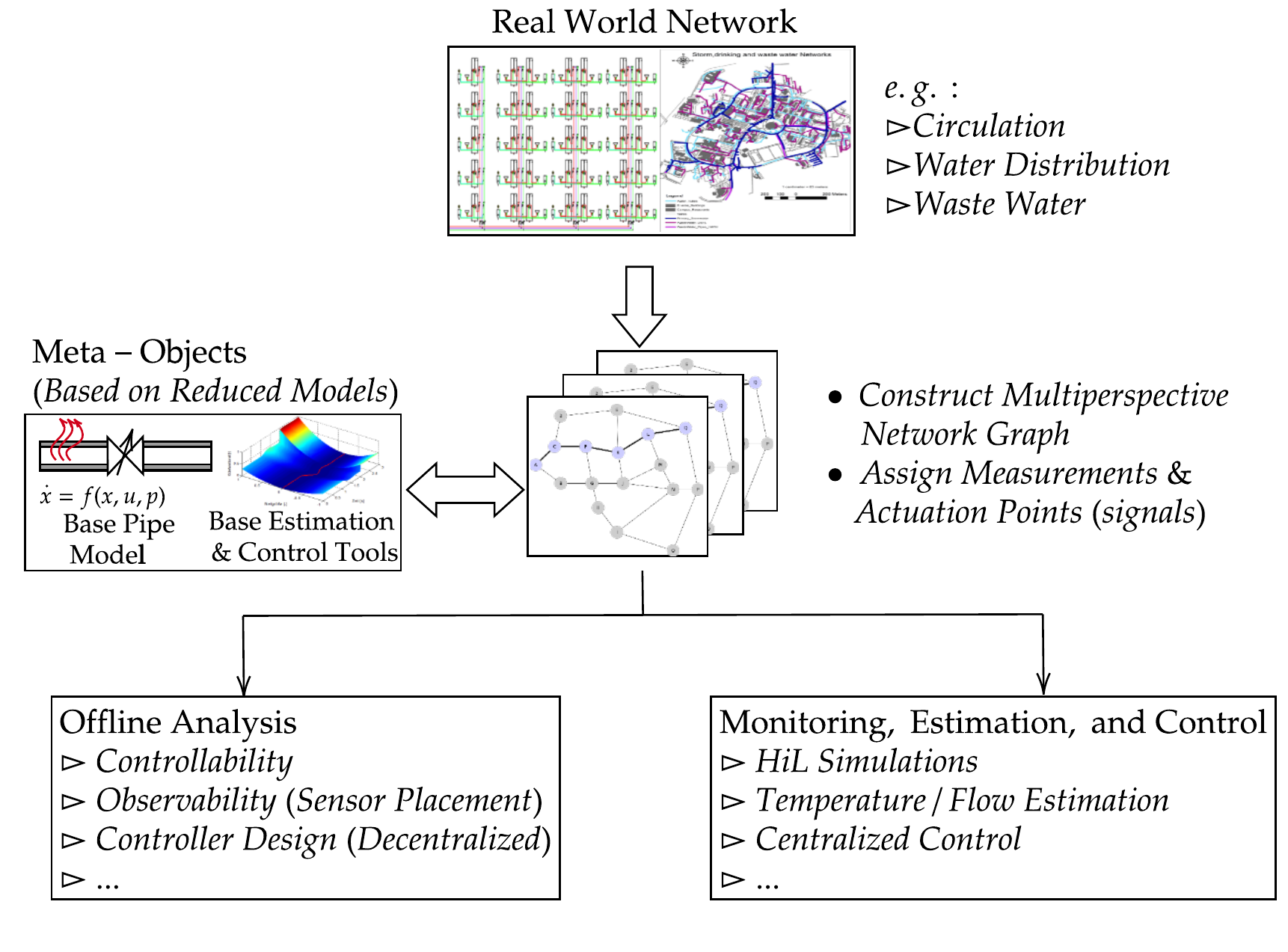}
	\vspace{-1cm}
	\caption{Schematic conceptualizing a holistic tool for water networks}
	\vspace{-0.5cm}
	\label{fig:LongTermVis}
\end{figure}

Water networks appear in a wide spectrum of applications ranging from domestic infrastructure to industrial plants. The task of monitoring and regulation of the water states within such networks is shared among most of these applications. The work at hand proposes and investigates a possible framework for the realization of a holistic tool for model-based analysis, monitoring and control of fluid networks of different scales all while considering the spatial variation of the controlled quantity. A schematic of the proposed tool is shown in \reffig{fig:LongTermVis}. The general idea is to construct multiple models of the network considering several aspects, e.g. thermal, chemical/disinfectant concentration, biological/microbial development. This allows the spatio-temporal monitoring of theses several water aspects for the modeled network. Moreover, the different actuation and sensing points within the network can be mapped to the model. Consequently, several model-based analysis and synthesis tasks can be performed, e.g. optimal sensor placement, controller design. One important requirement for the models used is their real-time capability for the monitoring and control tasks.
On the other hand, the local elements constituting the network should preserve their accuracy regarding the spatial variation of the fluid property under consideration. That is to say that the distributed parameter character of the models is essential and can not be neglected. This emphasizes the two contradicting incentives of high fidelity modeling and computational complexity reduction required for real-time purposes. Open source as well as commercial tools representing water networks efficiently widely used for network analysis and monitoring (specially hydraulic) tasks exist, see \cite{Review_Tools} for a review. The most popular tool, which most of the tools are based on, can be claimed to be EPANET \cite{rossman1999epanet} developed by the U.S. Environmental Protection Agency which also include extensions integrating other water aspects such as water quality \cite{WQ_Epanet} and extensions allowing for real-time tasks, e.g. \cite{RTP_Epanet,RTP_Epanet_1}. Tools preserving the spatial resolution while pursuing real-time capability for the given tasks are to the authors knowledge still generally absent in literature. Most of the existing tools either dismiss the accurate representation of spatial variation of the modeled water properties or the model requirements needed for real-time model based monitoring and control, i.e. control oriented models. Hence, the dominating real-time control methods for water networks have been classical model free methods or methods relying on spatially lumped static models \cite{RTC_Review}. Recent efforts for considering full spatial resolution of the networks (considering water quality aspect) within a control oriented model can be also observed \cite{WQModels4MPC}, where also a complexity reduction method for the constructed monolithic models are proposed \cite{WQMOR}. However, the general strategy used in \cite{WQModels4MPC,WQMOR} of formulating a monolithic model (emerging from direct spatial and temporal discretization of the whole network) and reducing the network's model as a whole is unsuitable for the goals of the proposed tool here. As the reduced model in such case loses its structural interpretability, where it is impossible to trace the network local elements to certain part of the reduced order model. Moreover, the network parameters are absorbed within the reduced model indicating that the model could not be used for tasks requiring variation of the parameters, e.g. parameter identification. Hence, the paradigm shift of using reduced order meta-models for each local part of the network is adopted here. It should be clear to this point that investigating and developing reduction methods allowing for preserving the model accuracy with a reduced computational cost is essential to guarantee the practicability of the proposed tool. Moreover, the reduced models should be optimized reaching the lowest possible computational complexity enhancing the capability of the tool to construct networks of a different scales. Model Order Reduction (MOR) techniques offer solutions to the problem allowing for the reduction of computational complexity. Most importantly, in order to utilize the Reduced Order Models (ROMs) for identification and meta-modeling purposes, the model's parametric dependencies must be preserved throughout the reduction process. This allows ROMs representing the network local element to be complemented within and employed by the other analysis and control modules of the tool, e.g. for parameter identification. Literature is rich with projection based MOR methods for finite dimensional systems. However, most of the methods developed for nonlinear systems have an empirical nature heavily relying on training data, and hence with no guarantee for parametric dependency and structure preservation. For example, in Proper Orthogonal Decomposition (POD) \cite{MORParamOverview,POD_MOR}, Full Order Model (FOM) simulations are directly used for the computation of the ROM basis by applying principal component analysis on the so-called snapshots matrix containing indexed spatio-temporal evolution of the FOM. Hence finding the spatial modes with the highest contribution to the dynamics and neglecting the rest. Alternatively in Trajectory Piecewise Linear Approximation (TPWL) methods \cite{TPWL_M}, the ROM is formulated as a collection (weighted sum) of local linear models derived from FOM simulation data. On the other hand methods developed for linear systems overcome such reliance on training data establishing rigor regarding error bounds, structure preservation and even ROM error optimality (see \cite{MORLin,MOROverview1,MOROverview2}). Some of these methods such as balanced truncation or moment matching methods with their advantages are being recently transferred for systems with weak/structural nonlinearity such as bilinear \cite{MORbilinH2Zhang.2002}, quadratic \cite{MORQb} and polynomial \cite{MORpoly} systems. Developing globally optimal reduction methods for general nonlinear systems can be claimed to be an intractable task. Therefore, the main strategy adopted in this work for addressing ROM optimality and structure preservation is by focusing on the model formulation. That is to allow a weak nonlinear form for which optimal reduction methods are feasible.

This contribution represents a first trial for the realization of the models used in the proposed tool (\reffig{fig:LongTermVis}) also evaluated on a real test bench of an exemplary circulation network. Firstly, a first-principles model for the thermal aspect of the flow of a single pipe presented and validated in \cite{Elkhashap.2021b} is briefly summarized along with the calculation scheme for the pipe hydraulic states. Afterwards, a novel Reduced ROM formulation relying on casting the system into standard bilinear form is presented. Moreover, the parameter dependencies are decoupled from the system matrices using matrix expansions decoupling the ROM matrices construction from the evaluation of the parameters. Hence, the ROM can be directly evaluated at any chosen parameter value. This allows the usage of the single pipe model as a meta-object representing all the constituting parts of the constructed network given correct parameterization of each. The proposed framework is then applied to the exemplary network present at the company VIEGA composed of a 60 $[m]$ water circulation network of three paths each of which including a demand point. 

The paper is organised as following. First, the single pipe meta-model is introduced, along with it's proposed parametric MOR technique. Second, the proposed global network flow solver for the tool is presented highlighting the main implementation novelty. Third, the experimental test bench under consideration is presented briefly mentioning the identification experiments and illustrating the validation experiments performed. Finally, the validation experiments results, i.e. single pipe HiL and total network model, are presented followed by a brief conclusion and outlook.

\section{single pipe meta-model}
The model constructed in this section is used as a building block for the construction of water network models with arbitrary topology and number of elements. The main focus in this contribution is upon the thermal dynamics of the fluid. However, the same methodology can be considered to describe other aspects such as water quality, e.g. microbial development, disinfectant concentration \cite{Elkhashap2022model}.


\subsection{Single Pipe Distributed Parameter Thermal Model}
\begin{figure}[h!]
	\centering
	\resizebox{\linewidth}{!}{
	\input{Figures/Pipe.tex}
	}
	\caption{Schematic of the single pipe considered for modelling}
	\vspace{-0.2cm}
	\label{fig:pipe}
\end{figure}
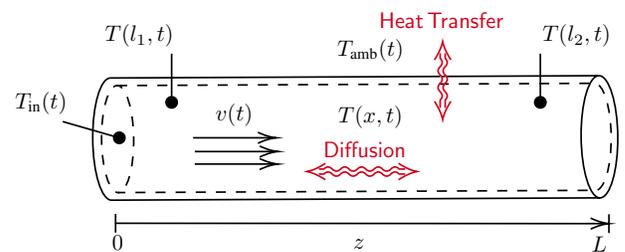

A schematic showing an abstracted pipe element considered for modeling is shown in \reffig{fig:pipe}.

The temperature $\Tf$ dynamics of water across a pipe can be modeled using the following second-order PDE in one spatial dimension $\Svar\in[0,\,L]$ defined over the length $L$ of the pipe
 \begin{subequations}
	\begin{equation}
	\dfrac{\partial{ \Tf}}{\partial t} + v(t) \dfrac{\partial \Tf }{\partial \Svar} = -\lambda(\Tf -T_{\mathrm{amb}}(t)) + D\dfrac{\partial^2 \Tf }{\partial \Svar^2},
	\label{eq:PDE}
	\end{equation}
	\begin{equation}
	\Tf(0, t)=T_{\mathrm{in}}(t),\quad {\dfrac{\partial \Tf}{\partial \Svar}}{\bigg|}_{\Svar=L}=0\label{eq:PDE_BC},
	\end{equation}
	\label{eq:PDEall}
\end{subequations}
with the flow velocity $v(t)$, effective heat transfer coefficient $\lambda$, the axial diffusion coefficient $D$, and the ambient temperature $T_{\mathrm{amb}}(t)$. For more details on the model derivation and validation the reader is referred to \cite{Elkhashap.2021b}. The flow velocity $v(t)$ is calculated using the volumetric flow rate $q(t)$ and the pipe cross section area $a$ assumed uniform along the pipe length, i.e. $v(t)=q(t)/a$. In case of a pipe with non uniform cross section area, an effective cross section area can be identified with the model parameters and used \cite{Elkhashap.2021b}, or the pipe can be split and represented by several model instances.
The PDE (\ref{eq:PDEall}) is solved with Method of Lines (MoL) \cite{Hamdi.2007}, where a semi-discretization using a uniform spatial grid of $\np$ points and a segment length of $\Delta \Svar=\frac{L}{\np}$ is constructed, i.e. $T(i\Delta \Svar,t)=T_i(t),\,\forall i \in \{0,...,\np\}$. The spatial partial derivatives are calculated on the grid using finite differences, where the convective term is approximated using first order upwind scheme and the diffusive term using a second order central scheme. This semi-discretization of the PDE including the boundary condition produces a nonlinear differential algebraic system (DAE). The nonlinearity of the emergent system have a special bilinear structure, mainly due to the appearance of the product between the PDE parameters, e.g. $v(t)$, and the temperature $\Tf(\Svar,t)$. The algebraic constraint emerges from the inlet boundary condition, i.e. $T_{\text {0}}(t)=T(0,t)$.  However, algebraic constraints introduce complications when it comes to MOR or to the integration of the state space equation. Hence, the constraint is eliminated by a ghost point on the left boundary \cite{Abgrall.2017}, i.e. excluding $T_{\text {0}}(t)=T(0,t)$ from the state space vector, while including the extra virtual input $v(t)T_{\mathrm{in}}(t)$ in the input vector. Consequently, the bilinear form is preserved and the first equation of the state space can be constructed. Hence finally, defining the state space vector $\Svec \in \mathbb{R}^{\np}$, the input vector $\Uvec\in \mathbb{R}^{4}$, and parameter vector $\Pvec\in \mathbb{R}^{3}$
\begin{align}
    &\Svec:=( T_1,\,\cdots,\,T_\np )^T\\
    &\Uvec:=( v(t),\,T_{\mathrm{in}}(t),\,v(t)T_{\mathrm{in}}(t),\,T_{\mathrm{amb}}(t) )^T,\\
    &\Pvec:=( \lambda,\,D,\, \Delta z)^T
\end{align}
produces the bilinear system
\begin{subequations}
\begin{equation}
    \dot{\Svec}=\matr{A}(\Pvec)\Svec+\sum_{i=1}^{4}\matr{Q}_i u_i \Svec +\matr{B}(\Pvec)\Uvec,\label{eq:FOM}
\end{equation}
\begin{equation}
\vek{y}=\matr{C}\Svec\label{eq:FOM_out}    
\end{equation}\label{eq:FOM_Tot}
\end{subequations}
with the system matrix $\matr{A} \in \mathbb{R}^{\np \times \np}$, input matrix $\matr{B}\in \mathbb{R}^{\np\times\udim}$, and measurement matrix $\matr{C} \in \mathbb{R}^{\ydim \times\np}$. The matrices $\matr{Q}_i \in \mathbb{R}^{\np\times\np}, \forall i\in\{1,..,\udim\}$ are the frontal slices of a $3^{rd}$ order tensor \cite{TensorDec} $\matr{\mathcal{Q}} \in\mathbb{R}^{\np\times\np\times\udim}$ describing the bilinear terms propagating in the system. This bilinear term can be also expressed using the Kronecker product notation $\otimes$ as 
\begin{equation}
\sum_{i=1}^{\udim}\matr{Q}_iu_i\Svec=\Qf\Uvec\otimes\Svec \label{eq:GenBilin}
\end{equation}
where $\matr{\mathcal{Q}}^{[1]}=[\matr{Q}_1,\cdots,\matr{Q}_m]\in \mathbb{R}^{n\times nm}$ is the mode-1 matricization \cite{TensorDec} of the tensor $\matr{\mathcal{Q}}$. This notation will be useful later for parametric dependency decoupling as well as the compact representation of the ROM. The FOM matrices with the above illustrated solution method are then
\begin{subequations}
\begin{equation}
\matr{A}=\begin{bmatrix}
\Theta   &\beta   &\ldots&0\\
\beta   & \Theta  &\ddots&\vdots\\
\vdots&\ddots&\ddots   &\beta\\
0     &\ldots&\beta   &\Theta
\end{bmatrix},\quad
\matr{B}=\begin{bmatrix}
0     &\beta &\theta&\lambda\\
0     &0     &0&\lambda\\
\vdots&\vdots&\vdots&\vdots\\
0     &0     &0&\lambda\\
\end{bmatrix},
\label{eq:FOMmat} 
\end{equation}
\begin{equation}
\matr{Q}_1=\frac{1}{\Delta \Svar}\begin{bmatrix}
1&0 &\ldots&0\\
-1& \ddots&\ddots&\vdots\\
\vdots&\ddots&1&0\\
0&\ldots  &-1 &1
\end{bmatrix},\quad
\end{equation}
\begin{equation}
\matr{Q}_2=\matr{Q}_3=\matr{Q}_4=\matr{\bar{0}_{n\times n}},\quad 
\end{equation}
\begin{equation}
 \Theta=-\lambda-\frac{2D}{\Delta \Svar^2},\quad \beta=\frac{D}{\Delta \Svar^2}.   
\end{equation}
\end{subequations}
The measurement matrix $\matr{C} \in \mathbb{R}^{n_m\times\np}$ has number of rows corresponding to the number of available temperature measurements $n_m$ along the pipe. The matrix is populated with ones placed at the position described by the measurement/sensor number ($i^{th}$ row) and its corresponding spatial position ($j^{th}$ column) 
\begin{equation}
c_{ij}=\begin{cases}1 & \text{sensor \ensuremath{i} available within } j\Delta \Svar<\Svar < (j+1)\Delta \Svar \\ 0 & \text{otherwise} 
\end{cases}
\label{eq:C}
\end{equation}
\subsection{Model Order Reduction}
The model introduced in the previous section \refeq{eq:FOM_Tot} fulfills one of the main goals of the tool, which is the predictive spatial resolution of the model. The fluid temperature at any spatial point - on the defined grid - across the pipe can be retrieved and hence monitored or controlled. However, this implicates a differential system \refeq{eq:FOM} with a very high dimension $\np$, which compromises the real-time capability of the model. Therefore, projection-based MOR is employed in order to derive a model with low computational complexity. Projection-based MOR methods \cite{MOROverview1}  general assumption is that the state evolution of the $\np$ high dimensional differential does not exploit the whole state space degrees of freedom but usually utilizes a $\nr$ dimension subspace with $\nr\ll \np$ . Hence the complete information of the state evolution can be expressed approximately in terms of this subspace using suitable projection matrices
\begin{equation}\label{eq:ROMAssum}
\vek{x}(t)\approx\vek{V}\vek{x}_r(t), \quad \vek{x}_r\in\mathbb{R}^{\nr},\quad \nr\ll \np,
\end{equation}
with the new reduced order state vector $\vek{x}_r$ and the projection matrix $\vek V\in \mathbb{R}^{\np\times \nr},\,\nr=\text{rank}(\vek V)$ consisting of the basis vectors of this lower order subspace (i.e. trial basis). Applying the Petrov-Galerkin condition \cite{MORParamOverview}, where the approximation residual $\vek{V}\vek{\dot{x}}_r-\vek{f}(\vek{V}\vek{x}_r,\Uvec,\Pvec)$, with $\vek{f}(\cdot)$ being the right hand side of \refeq{eq:FOM}, is forced to vanish by introducing the matrix $\vek{W}\in\mathbb{R}^{\np\times\nr},\,\nr=\text{rank}(\vek{W})$ (i.e. test basis). This produces the following reduced order differential equivalent of \refeq{eq:FOM}
\begin{subequations}
\begin{equation}
\vek{{\dot{x}}_r}=\Ar\vek{x}_r+\Qr \vek{u}\otimes \vek{x}_r + \Br \vek{u},\label{eq:ROM}
\end{equation}	
\begin{equation}
\vek{y}=\Cr \vek{x}_r\label{eq:ROM_out}
\end{equation}
\end{subequations}
with the reduced order system matrices $\Ar \in \mathbb{R}^{\nr\times \nr}$, $\Qr \in \mathbb{R}^{\nr \times \udim\nr}$, $\Br \in \mathbb{R}^{\nr \times \udim}$, and $\vek{\hat{C}} \in \mathbb{R}^{l\times \nr}$ as following
\begin{align}
&\Ar=\vek{T}\vek{A}(\Pvec)\vek{V},\quad \Br=\vek{T}\vek{B}(\Pvec),\quad\Cr=\vek{C}\vek{V},\\
&\Qr=\Qrf,\quad \vek{T}=(\vek{W}^T\vek{V})^{-1}\vek{W}^T.
\end{align}

Now for the case at hand, there are three main requirements for the MOR procedure employed :
\begin{enumerate}[label=(\alph*)]
    \item minimal prediction error due to reduction,
    \item structure and parametric dependency preservation,
    \item significant computational complexity reduction,
\end{enumerate}
In order to address the first requirement, the $\mathcal{H}_{\mathrm{2}}$ norm method for bilinear systems \cite{MORbilinH2Zhang.2002,MORbilinH2,Elkhashap.2021a} is employed utilizing the bilinear form. This method constructs the optimality conditions of the error system on the $\mathcal{H}_{\mathrm{2}}$ norm, which is expressed via the reachability and observability Gramians derived from Volterra series representation of bilinear systems. These conditions are shown to satisfy a certain pair of generalized Lyapunov equations \cite{MORbilinH2Zhang.2002}. The optimal reduced system is found by solving a constrained minimization problem in \cite{MORbilinH2Zhang.2002}. However, in \cite{MORbilinH2} two iterative algorithms are proposed for the determination of the basis leading to the minimal $\mathcal{H}_{\mathrm{2}}$ norm error system of which the first algorithm \cite[p.~14]{MORbilinH2}, \cite{Elkhashap.2021a} is used here. In order to address requirement (b) of the preservation of the parametric dependency in the ROM, a matrix expansion based decoupling of the parameter is performed. Defining a the vector field $\vek{g}(\Pvec) \in \mathbb{R}^2$
\begin{equation}
\vek{g}(\Pvec) =[g_1(\Pvec),\,g_2(\Pvec)]=[\lambda,\,D/\Delta \Svar^2]^T,    
\end{equation}
the matrix $\matr{A}(\Pvec)$ can be decoupled from the parameter vector $\Pvec$ as follows 
\begin{subequations}
\begin{equation}
\matr{A}(\Pvec)=\lambda\Amat_1+\frac{D}{\Delta \Svar^2}\Amat_2=\sum_{i=1}^{2}g_i(\vek{p})\Amat_i
\end{equation}
\begin{equation}
\Amat_1=-\Imat{\np},\quad
\Amat_2=\begin{bmatrix}
-2   &1   &\ldots&0\\
1   & -2  &\ddots&\vdots\\
\vdots&\ddots&\ddots   &1\\
0     &\ldots&1   &-2
\end{bmatrix},
\end{equation}
\end{subequations}
now with $\At=[\matr{A}_1,\, \matr{A}_2]$ the parameter-independent matricized tensor $\Atr$ can be calculated 
\begin{equation}
    \Ar\vek{x}_r=\matr{T}\matr{A}(\Pvec)\matr{V}\vek{x}_r=\underbrace{\matr{T}\At(\Imat{2}\otimes\matr{V})}_{\large\Atr}(\vek{g}(\Pvec)\otimes\vek{x}_r). 
\end{equation}
The input matrix $\Br$ can be decoupled from the parameter vector $\Pvec$ analogously with the suitable mapping $\vek{h}(\Pvec)$:
\begin{equation}
    \Br\vek{u}=\matr{T}\matr{B}(\Pvec)\vek{u}=\underbrace{\matr{T}\Bt(\Imat{3}\otimes\matr{V})}_{\large\Btr}(\vek{h}(\Pvec)\otimes\vek{x}_r), 
\end{equation}
Hence, finally the parameter-independent ROM matrices $\Atr,\,\Btr$ are calculated using $\vek{V},\,\vek{W}$ from the above mentioned $\mathcal{H}_{\mathrm{2}}$ norm method delivering the following form of of the reduced ODE
\begin{equation}
\vek{{\dot{x}}_r}=\Atr\vek{g}(\Pvec)\otimes\vek{x}_r+\Qr \vek{u}\otimes \vek{x}_r + \Btr \vek{h}(\Pvec)\otimes\vek{u}.\label{eq:ROMfin}
\end{equation}	
Note that the expansion used for decoupling the parameter from the system matrices introduces extra computational burden. This mainly due to the expansion of the system matrices used in projection, e.g $\Atr \in \mathbb{R}^{r\times2r}$ instead of $\Ar \in \mathbb{R}^{r\times r}$, as well as due to the evaluation of the parameter mappings $\vek{g}(\Pvec)$ and $\vek{h}(\Pvec)$. Nevertheless, this increase in the computational burden is mainly within the offline calculations of the ROM matrices with no increase at all for the online calculations. This is because the parameter vector usually would not be varied online and a fixed ROM matrices evaluated at a certain parameter values $\Pvec^*$ are calculated offline
\begin{equation*}
   \Atr\vek{g}(\Pvec^*)\otimes\vek{x}_r=\underbrace{\Atr(\vek{g}(\Pvec^*)\otimes\Imat{r})}_{\large\Ar(\Pvec^*)}\vek{x}_r.
\end{equation*}
Nevertheless, in case of for systems with time varying parameters such evaluation is computationally cheap as all of the calculation involved matrices are of the reduced dimension $r$. Finally, in order to ensure requirement (c), the integration of the ROM ODE, i.e. \refeq{eq:ROM}, is implemented using the tool CasADi \cite{Andersson2019} exploiting algorithmic differentiation and C-code generation. Moreover, the natural sparsity in the FOM matrices, also the expanded matrices, is exploited utilizing sparse linear algebra routines, hence further increasing the computational efficiency of the calculations. Now after achieving a ROM with the parametric dependencies preserved, it can be used to represent the several elements of a network after adapting the parameters according to each element configuration. Moreover, it can be directly utilized for parameter identification purpose. 
\subsection{Single Pipe Hydraulic Model}
Since water is mainly an incompressible fluid, the transient hydrodynamic aspect of water flow in pipes is mostly non influential to other properties of interest, e.g. thermal, water quality\footnote{Nevertheless, the same framework can be adopted to analyse the spatially varying pressure dynamics in water pipes, e.g. for water hammers phenomena analysis by constructing distributed parameter model of using the water pressure as property.}. Consequently, a simplifying assumption of instantaneous change of the flow velocity within a pipe according to the difference of hydraulic pressure across the pipe is made. The relation between the pressure drop across the pipe $\Delta h$ and the volumetric flow rate $q(t)$  within the pipe is \cite{Elkhashap.2021b}
\begin{subequations}
\begin{equation}
    \Delta h= r_{\text {p}} q(t)^2
\end{equation}
\begin{equation}
r_{\text {p}}=\underbrace{\frac{\rho}{2a^2}f(\Rey,\epsilon)\frac{L}{d}}_{\text{friction resistance}}  +\underbrace{\frac{\rho}{2a^2}k_{\mathrm{min}}}_{\text{minor  losses}}+\underbrace{\rho (\frac{1}{c_{\text d}a_{\text v}(t)})^2}_{\text{valve resistance}}
\end{equation}
\end{subequations}
with the total pipe flow resistance $r_{\text {p}}$, fluid density $\rho$, pipe cross section $a$, pipe effective inner diameter $d$, minor losses dimensionless coefficient $k_{\mathrm{min}}$, and finally valve discharge coefficient $c_{\text d}$. The Darcy friction coefficient $f(\Rey,\epsilon)$ is dependent on the flow regime, i.e. Reynolds number $\Rey$, and the pipe roughness $\epsilon$ (in case of turbulent flow). The following calculation model for $f$ is adopted
\begin{itemize}
		\item if $\Rey<2400$ (Laminar Flow): $f= 64/\Rey$,
		\item if $\Rey>2400$ (Turbulent Flow) : solve Colebrook White implicit correlation with initialization using Haaland explicit relation \cite{pimenta2018performance}.
\end{itemize}
The pipe valve opened area $a_{\text v}(t)$ is represented as simple linear mapping of the valve actuation signal $u_{\text v}(t) \in [0,\,1]$ using the valve fully closed $a_{\text v}^{fc}$ and fully  opened area $a_{\text v}^{fo}$
\begin{equation*}
    a_{\text v}(t)=a_{\text v}^{fc}+u_{\text v}(t)(a_{\text v}^{fo}-a_{\text v}^{fc}).
\end{equation*}

\section{Full Network Model}
In order to predict the fluid temperature of the whole network using the ROM representing the local elements in the network, the flow rates of all of the elements constituting the network must be first estimated. Hence, a global network flow solver determining the hydraulic states (pressure at nodes and flow at links) is necessary. Moreover, models representing the hydraulic flow resistance for each pipe as well as models representing the discharge flow (demand flow) at nodes are needed. Consequently, the single pipe hydraulic model represented in the previous section is employed. The demand flow at nodes is furthermore modeled similar to the valve model within the single pipe hydraulic model by introducing demand valves control signal $u^d(t)$ and a constant emitter discharge coefficient $c_{\text d}^e$ for all demand points.
The network flow problem is solved using Global Gradient Algorith (GGA) \cite{rossman1999epanet,todini1988gradient,GGA2011} which applies Newton-Raphson technique for solving the nonlinear matrix problem emerging from balancing the heads and the flows of the whole network simultaneously. An explicit form of Newton-Raphson iterations can be derived, consequently the problem solver iterations are reduced to a set of linear algebraic evaluations. The method has the advantage of fast convergence, also the initial (trial) flow rate values must not satisfy the continuity (mass conservation) equation to achieve convergence. For a network with $n_{\text {p}}$ pipes, $n_{\text {n}}$ unknown nodal pressures, and $n_{\text {0}}$ known nodal pressures \footnote{The knowledge of at the least one head (pressure) at any node in the network is a requirement for the existence of a unique solution for the network flow problem} (e.g. reservoir, head pump), the steps undertaken for the formulation of the network to generate an efficient tailored hydraulic solver for the network using GGA can be summarized into the following steps:
\begin{enumerate}
    \item Construct and visualise the network.
    \item Assign actuation, i.e. valves, and measurement signals (pressure, temperature).
    \item Assign demand points.
	\item Transform the network into a directed graph.
	\item Define the vectors required for the matrix problem formulation
	\begin{align*}
	&\text{pipes flow vector}:\vek{q}_{\text {p}}\in \mathbb{R}^{n_{\text {p}}},\\
	&\text{pipes flow resistance vector}:\vek{r}_{\text {p}}(\vek{q}_{\text {p}})\in \mathbb{R}^{n_{\text {p}}},\\
	&\text{unknown nodal pressure vector}:\vek{h}_{\text {n}}\in \mathbb{R}^{n_{\text {n}}},\\
	&\text{nodal demand flow vector}:\vek{q}_{\text {n}}\in \mathbb{R}^{n_{\text {n}}},\\
	&\text{known nodal pressure vector}:\vek{h}_{\text {0}}\in \mathbb{R}^{n_{\text {0}}}.
	\end{align*}

	\item Determine the network directed graph and the corresponding incidence matrix $\mathbf{\bar{A}}_{\text {pn}} \in \mathbb{R}^{n_{\text {p}}\times(n_{\text {0}}+n_{\text {n}})}$.
	\item Extract the topological matrices \cite{GGA2011} $\mathbf{A}_{\text {p0}}\in \mathbb{R}^{n_{\text {p}}\times n_{\text {0}}}$ and $\mathbf{A}_{\text {pn}}=\mathbf{A}_{\text {np}}^T \in \mathbb{R}^{n_{\text {n}}\times n_{\text {p}}}$, which are sub-matrices of the incidence matrix $\mathbf{\bar{A}}_{\text {pn}}$ according to the known nodal pressures in the network.
	\item Construct the diagonal matrix
	\begin{equation*}
	    \mathbf{A}_{\text {pp}}=\mathrm{diag}(\vek{r}_{\text {p}})\mathrm{diag}(\abs{\vek{q}_{\text {p}}})=\begin{bmatrix}
		r_1|q_1|   & \dots & 0\\
		\vdots     & \ddots& 0\\
		0          & \dots & r_{\text {p}}|q_{\text {p}}|\\	
		\end{bmatrix}.
	\end{equation*}
	\item Now finally the network global mass and energy balance is summarized into the following matrix equation \cite{GGA2011}:
	\begin{align}\label{eq:NREqu}
		\begin{bmatrix}
		\mathbf{A}_{\text {pp}} &\mathbf{A}_{\text {pn}}\\
		\mathbf{A}_{\text {np}} & \matr{0} \\
		\end{bmatrix}
		\begin{bmatrix}
		\vek{q}_{\text {p}}\\
		\vek{h}_{\text {n}}\\
		\end{bmatrix}-
		\begin{bmatrix}
		\mathbf{A}_{\text {p0}}\vek{h}_{\text {0}}\\
		\vek{q}_{\text {n}}\\
		\end{bmatrix}=\matr{0}.
	\end{align}
	\item Derive the gradient-data matrices, i.e. required jacobians
	\begin{equation}
	\mathbf{D}_{\text {pp}}=\dfrac{\partial (\mathbf{A}_{\text {pp}}\vek{q}_{\text {p}})}{\partial\vek{q}_{\text {p}}},\quad \mathbf{D}_{\text {nn}}=\dfrac{\partial \vek{(q}_{\text {n}})}{\partial\vek{h}_{\text {n}}}    ,
	\end{equation}
	
	\item Use the derived matrices for the generation of explicit expressions of the Newton-Raphson iteration solving for $\vek{q}_{\text {p}}$ and $\vek{h}_{\text {n}}$ \cite{GGA2011,todini1988gradient}.
\end{enumerate} 
The construction and visualization of the networks is performed using \textsc{MATLAB} \cite{MATLAB_2020}, where an automated procedure is constructed for the generation of the network directed graph and related matrices, i.e. step 1-4). Moreover, utilizing \textsc{MATLAB}'s symbolic toolbox, the required jacobians are calculated symbolically generating a \textsc{MATLAB} function for the explicit evaluation of the Newton-Raphson iteration tailored for the defined network, i.e. step 5-8). The initialization of the iterations can be arbitrarily chosen. However, shifting the previous time step solution to be the initial value for the next time step is performed to improve the convergence rate. Finally, the distributed parameter model of the water temperature is used to represent each link in the directed graph. The network incidence matrix $\mathbf{\bar{A}}_{\text {pn}}$ is used to map the temperature predictions to the correct node. A simple weighted average (according to the flow ratio) is used for the temperature at nodes where fluid mixing occurs.

\section{Experiments}
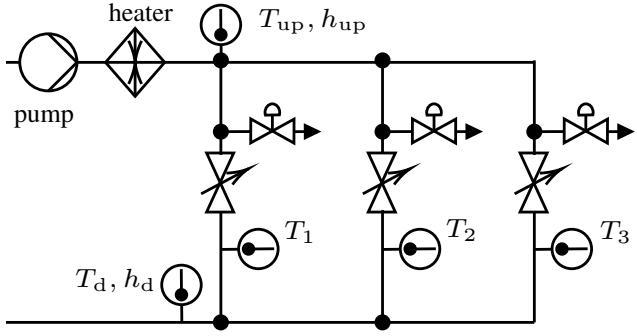
\begin{figure*}[t!]
	\centering
	\resizebox{\textwidth}{!}{
	\input{Figures/Test_Bench_DG.tex}
	}
	\vspace{-0.6cm}
	\caption{Schematic of the experimental test bench (left) and the corresponding directed graph of the test bench (right)}
	\vspace{-0.5cm}
	\label{fig:TB_DG}
\end{figure*}
The methods introduced in the previous sections are applied to a $60\,[m]$ experimental test bench present at VIEGA GmbH \& Co. KG, Attendorn, Germany. The test bench includes three flow paths where each path equipped with a metered discharge valve at the middle representing the demand points. Moreover, a flow control valve is installed within each path with which the flow configuration in the different paths can be varied. Finally, water temperature as well as pressure sensors delivers measurements for sparse spatial points along each path of the test bench. A simplified schematic as well as the corresponding directed graph of the experimental test bench are shown in \reffig{fig:TB_DG}. For more details on the test bench construction, actuation, and sensor setup the reader is referred to \cite{Elkhashap.2021b}. Several validation scenarios have been performed, However, only the following scenarios will be presented hereafter
\begin{enumerate}[label=(\alph*)]
    \item single flow path, i.e. with path $1$ and $3$ valves completely closed, Hardware in the Loop (HiL) validation using the ROM,
    \item full network model validation.
\end{enumerate}
As the pipe configuration of the three paths are very similar, only one path is used for the identification of the thermal parameters $\lambda,\, D$ regarded as time-invariant and constant for the three paths. Hence, the pipe thermal parameters identified from scenario (a) are applied to all of the pipe elements in scenario (b). A FOM with $\np=1200$, i.e. prediction spatial resolution of $50\,\mathrm{mm}$, is used where the corresponding ROM using the above illustrated method and reduced dimension $r=7$ is constructed. It should be noted that the FOM $\np=1200$ spatial discretization does not reflect a need for a certain spatial accuracy but rather to test the method for a challenging discretization case (examine the performance in case of a scale up). The ROM is then used directly in the parameter identification fitting the experimental data to the ROM predicted output. 

\subsection{Identification}

The test bench was used to generate two separate experimental data-sets (of different operating scenarios) which are then used for the identification and validation, respectively.
The thermal parameters identification problem of a single path is formulated as a Nonlinear Program (NLP) and solved using Matlab's Sequential Quadratic Programming (SQP) based solver. For more information on the formulation and solution method of the identification problem the reader is referred to \cite{Elkhashap.2021b} as the model fitting problem is identical to the one introduced in the paper but here with the ROM directly utilized instead of the FOM. The network hydraulic parameters related to minor losses, flow control and demand valves, i.e. $c_{\text d},\,c^e _{\text d},\,k_{\text {min}}$ for each path, are identified using an experimental data-set of the full network operation. The fitting problem is also formulated as a NLP and solved analogously. The identification experiments and results are omitted here for brevity illustrating only the validation experiments and results.

\subsection{Single Path ROM Validation Experiments}
\begin{figure}[b!]
  \centering
    \psfrag{18}{11}
    \includegraphics[width=\linewidth]{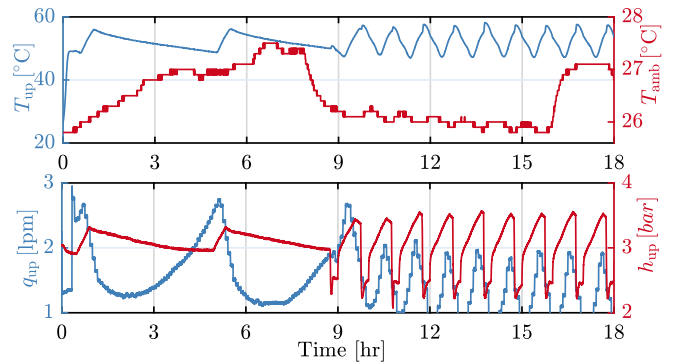}
    \vspace{-0.5cm}
  	\caption{HiL experiment input and disturbance profiles}
	\vspace{-0.2cm}
	\label{fig:HIL_IN}
\end{figure}

In order to evaluate the performance of the proposed models, i.e. ROM for temperature prediction and hydraulic model, several validation tests are run. Firstly, the models are validated against experimental data of one path of the test bench. Then an $18$ hour HiL experiment is performed. Presented here are only the results of the $18$ hour HiL validation experiment. The ROM is used to predict the downstream temperature of one of the fluid paths, i.e. the second, in real-time during a transient operation of the test bench. The generated ROM integrator code with a $1$ sec sampling period is deployed on the test bench operating PLC and left overnight (18 hours). The operating conditions of the 18 hour experiment, i.e. upstream temperature and pressure, ambient temperature, flow rate, are shown in \reffig{fig:HIL_IN}.

\subsection{Full Network Model Validation Experiment}

\begin{figure}[h!]
\vspace{-0.4cm}
  \centering
    \includegraphics[width=\linewidth]{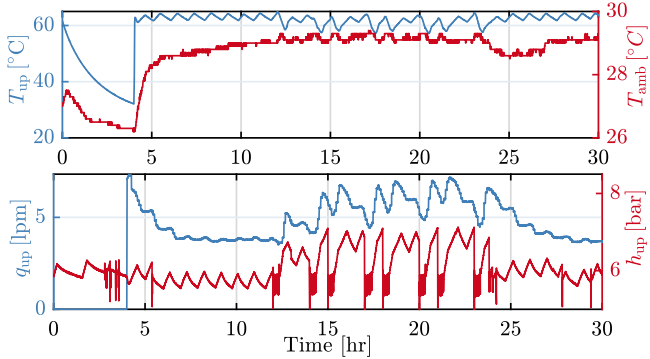}
    \vspace{-0.6cm}
  	\caption{Network model validation experiment input profiles}
	\vspace{-0.5cm}
	\label{fig:HIL_IN1}
\end{figure}

\begin{figure}[h!]
  \centering
    \includegraphics[width=0.97\linewidth]{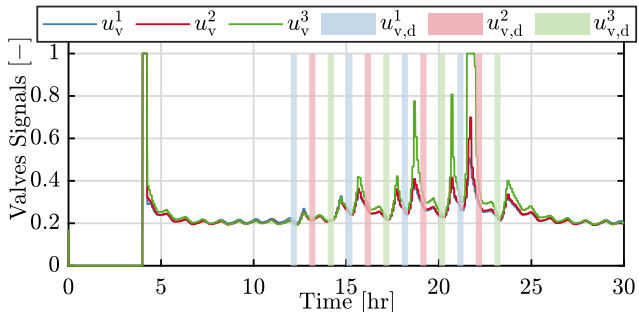}
    \vspace{-0.3cm}
    \caption{Flow control valves and demand valves (shaded area corresponding to fully opened) control signals used for the network validation experiment}
	\label{fig:HIL_IN2}
\end{figure}

In order to test the whole network model including the network flow solver, the full network model is validated in a separate experiment. The data-set generated for this validation comprises a 30 hour operation of all of the three paths of the test bench. The experiment contains a certain time period (first 5 hours) where the three control valves are fully closed in order to test a the ability of the ROMs to predict a dominating diffusive heat transfer phenomena (zero convective flow). Moreover, the three demand flow valves are activated several times for each path at different time points. The only known inputs to the network model are the network total feed $q_{\mathrm{up}}$, upstream pressure $h_{\mathrm{up}}=h_1$ and temperature $T_{\mathrm{up}}$, the ambient temperature $T_{\mathrm{amb}}$, and the path's flow control valves signals $u_{\text v}^1,u_{\text v}^2,u_{\text v}^3$ as well as the demand valves' control signals $u_{\mathrm{v,d}}^1,u_{\mathrm{v,d}}^2,u_{\mathrm{v,d}}^3$. The experiment input profiles as well as the valves signals are shown in \reffig{fig:HIL_IN1} and \ref{fig:HIL_IN2}, respectively.   

\section{Results}
\subsection{Single Pipe HiL Results}
\begin{figure}[t!]
  \centering
  \footnotesize
  \def\svgwidth{\linewidth}
  \vspace{-0.2cm}
  \includegraphics[width=\linewidth]{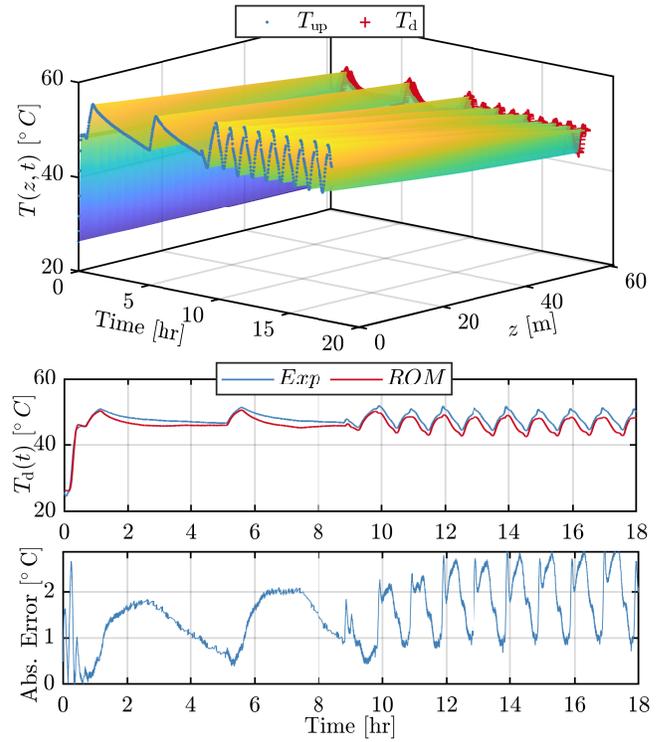}
  \vspace{-0.5cm}
  \caption{HiL experiment predicted temperature against measured temperature and the corresponding prediction error}
	\vspace{-0.2cm}
	\label{fig:HIL_R}
\end{figure}

The HiL simulation results are shown in \reffig{fig:HIL_R}. It can be observed that the ROM could predict down stream temperature accurately with an absolute error below $3\,\mathrm{^\circ C}$. The average computation time for the ROM one step prediction lied below 1 millisecond which shows the potential for the models exploitation into further real-time control applications.

\subsection{Full Network Validation}
\begin{figure}[t!]
  \centering
  \vspace{-0.1cm}
  \includegraphics[width=\linewidth]{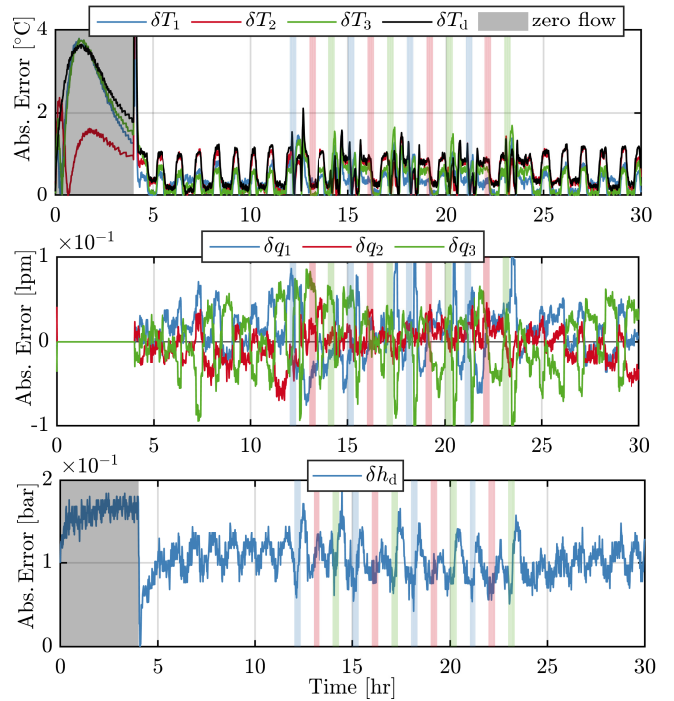}
  \vspace{-0.6cm}
  	\caption{Network model predictions absolute errors: temperatures (top), flow rates (middle), and downstream pressure (bottom)}
	\vspace{-0.5cm}
	\label{fig:Net_Err}
\end{figure}
The model predictions absolute errors with respect to the measured ground truth for the temperatures, flow rates as well as pressure are shown in \reffig{fig:Net_Err}. The network solver could estimate the flow rates as well as the downstream pressure effectively with maximum absolute error for the pressure and flow rates below $0.2 \,\mathrm{bar}$ and $0.1\,\mathrm{lpm}$, respectively (Mean Relative Error (MRE) below 2 \% for both). The local temperature ROMs could also predict the temperature evolution accurately with maximum absolute temperature error below $4\,\mathrm{^\circ C}$ (4 \% MRE), for the dominating diffusive case, i.e. zero flow velocity, and below $2\,\mathrm{^\circ C}$ (2 \% MRE) for the dominating convective flow case. However, the error for path $2$, the one used for parameter identification trials, stayed below $2\,\mathrm{^\circ C}$ for both cases. This indicates that a separate more detailed parameter identification for each path would increase the accuracy of the model (more specifically for the diffusion and heat transfer parameter). Furthermore, It can be concluded from the results that the time invariant assumption of the thermal parameters, i.e. $\lambda$ and $D$, causes the deviation. Hence, using time-variant parameters, e.g. using online estimation techniques, represents further improvement potential for the model's accuracy. The mean calculation time (on a Windows 10 development computer, Intel(R) Core(TM) i7-7700HQ CPU@2.8GHZ, 8GB RAM) of one step prediction of each ROM representing the network pipes is $2.26$ msec. Moreover, the mean calculation time of the network flow solver for a single time step is $12.5$ msec with the Newton iteration tolerance, i.e. for  \refeq{eq:NREqu}, set to $10^{-9}$.

\section{CONCLUSION}
A framework for modeling water networks is proposed with the preservation of spatial resolution of the property under consideration. A novel parameter-dependency preserving model order reduction method for the water temperature distributed parameter model is introduced. The ROM as well as the full network model are identified and validated in separate experiments showing the predictive accuracy as well as the real-time capability of the models. The water network model with predictive spatial resolution within millimeters could predict the water temperature evolution with mean relative error below 4 \%. The network hydraulic states, i.e. flow rates and pressures, are predicted with mean relative error below 2 \%. Furthermore, the model computation steps are within milliseconds order of magnitude emphasizing the real-time capability of the models, which is also demonstrated for a single path in a Hardware in the Loop experiment. For future work other water properties will be considered within the network model, more specifically the water quality, i.e. disinfectant concentration. Moreover, a centralized Nonlinear Model Predictive Controller (NMPC) is to be investigated for the temperature tracking at various node of the exemplary water circulation network.

\addtolength{\textheight}{-12cm}   






\bibliography{Paper_Main_IEEE_CP_Notice}
\bibliographystyle{IEEEtran}

\end{document}

%% file: MyCmd.tex
\newcommand{\refeq}[1]{(\ref{#1})}
\newcommand{\reffig}[1]{Fig.~\ref{#1}}

\usepackage{physics}
\newcommand{\matr}[1]{\boldsymbol{#1}}
\newcommand{\vek}[1]{\boldsymbol{#1}}

\DeclareSymbolFont{matha}{OML}{txmi}{m}{it}
\DeclareMathSymbol{\varv}{\mathord}{matha}{118}

\newcommand{\Rey}{\operatorname{\mathit{R\kern-.04em e}} }
\newcommand{\Pecl}{\operatorname{\mathit{P\kern-.08em e}}}
\newcommand{\Pran}{\operatorname{\mathit{P\kern-.03em r}}}
\newcommand{\Rayl}{\operatorname{\mathit{R\kern-.04em a}}}
\newcommand{\Nuss}{\operatorname{\mathit{N\kern-.09em u}}}
\newcommand{\Gras}{\operatorname{\mathit{G\kern-.05em r}}}

\renewcommand{\abs}[1]{\lvert#1\rvert}

\newcommand{\Tf}{T}
\newcommand{\Svar}{z}
\newcommand{\Svec}{\vek{x}}
\newcommand{\Uvec}{\vek{u}}
\newcommand{\Pvec}{\vek{p}}
\newcommand{\udim}{4}
\newcommand{\ydim}{l}
\newcommand{\np}{N}
\newcommand{\nr}{r}

\newcommand{\Imat}[1]{ \vek{I}_{#1 \times #1} }

\newcommand{\Ar}{ \vek{\hat{A}} }
\newcommand{\Qr}{  \vek{\mathcal{\hat{Q}}}^{[1]} }
\newcommand{\Qrf}{ \vek{T} \vek{\mathcal{Q}}^{[1]} (\vek{I}_{m\times m} \otimes \vek{V}) }
\newcommand{\At}{  \vek{\mathcal{A}}^{[1]} }
\newcommand{\Bt}{  \vek{\mathcal{B}}^{[1]} }
\newcommand{\Atr}{  \vek{\mathcal{\hat{A}}_r}^{[1]} }
\newcommand{\Btr}{  \vek{\mathcal{\hat{B}}_r}^{[1]} }

\newcommand{\Br}{ \vek{\hat{B}} }
\newcommand{\Cr}{ \vek{\hat{C}} }

\newcommand{\Amat}{\vek{A}}
\newcommand{\Qf}{ \vek{\mathcal{Q}}^{[1]} }

\usepackage{siunitx}
\newcommand{\refappendix}[1]{\hyperref[#1]{Appendix~\ref*{#1}}}

\usepackage{pgf}
\definecolor{lime}{HTML}{A6CE39}

\definecolor{lime}{HTML}{A6CE39}

\usepackage{psfrag}
\usepackage{pgf,tikz,pgfplots}
\pgfplotsset{compat=newest}
\usetikzlibrary{arrows,decorations.markings}
\usetikzlibrary{positioning}
\usepackage{tikz-network}
\usetikzlibrary{plotmarks}

\definecolor{mycolor1}{rgb}{0.25098,0.49804,0.71765}%

\usepackage[customcolors]{hf-tikz}
\usetikzlibrary{fit}
\tikzset{style green/.style={
		set fill color=green!20!lime!60,
		set border color=white,
	},
	style cyan/.style={
		set fill color=cyan!90!blue!60,
		set border color=white,
	},
	style orange/.style={
		set fill color=orange!80!red!60,
		set border color=white,
	},
	hor/.style={
		above left offset={-0.15,0.31},
		below right offset={0.15,-0.125},
		#1
	},
	ver/.style={
		above left offset={-0.1,0.3},
		below right offset={0.15,-0.15},
		#1
	}
}

\tikzset{%
	highlight/.style={rectangle,rounded corners,fill=red!80,fill opacity=0.3,thin,inner sep=0pt}
}

%% file: Figures/Pipe.tex
\tikzset{every picture/.style={line width=0.75pt}} 

\begin{tikzpicture}[x=0.75pt,y=0.75pt,yscale=-1,xscale=1]

\draw   (474.63,278.32) -- (183.01,279.7) .. controls (176.96,279.73) and (171.91,263.41) .. (171.74,243.25) .. controls (171.56,223.09) and (176.31,206.72) .. (182.36,206.69) -- (473.98,205.32) .. controls (480.02,205.29) and (485.07,221.61) .. (485.25,241.77) .. controls (485.43,261.93) and (480.67,278.29) .. (474.63,278.32) .. controls (468.58,278.35) and (463.53,262.03) .. (463.35,241.87) .. controls (463.17,221.71) and (467.93,205.35) .. (473.98,205.32) ;
\draw  [dash pattern={on 4.5pt off 4.5pt}] (186.32,211.22) -- (471.43,209.87) .. controls (476.72,209.85) and (481.14,224.14) .. (481.3,241.8) .. controls (481.46,259.46) and (477.29,273.8) .. (471.99,273.82) -- (186.89,275.16) .. controls (181.59,275.19) and (177.17,260.89) .. (177.01,243.24) .. controls (176.86,225.58) and (181.02,211.24) .. (186.32,211.22) .. controls (191.62,211.19) and (196.04,225.49) .. (196.2,243.15) .. controls (196.36,260.8) and (192.19,275.14) .. (186.89,275.16) ;
\draw    (185.37,293.5) -- (479.97,293.72) ;
\draw [shift={(479.97,293.72)}, rotate = 180.04] [color={rgb, 255:red, 0; green, 0; blue, 0 }  ][line width=0.75]    (0,5.59) -- (0,-5.59)(10.93,-3.29) .. controls (6.95,-1.4) and (3.31,-0.3) .. (0,0) .. controls (3.31,0.3) and (6.95,1.4) .. (10.93,3.29)   ;
\draw [shift={(185.37,293.5)}, rotate = 180.04] [color={rgb, 255:red, 0; green, 0; blue, 0 }  ][line width=0.75]    (0,5.59) -- (0,-5.59)   ;
\draw    (218.86,221.26) -- (218.86,192.13) ;
\draw [shift={(218.86,221.26)}, rotate = 270] [color={rgb, 255:red, 0; green, 0; blue, 0 }  ][fill={rgb, 255:red, 0; green, 0; blue, 0 }  ][line width=0.75]      (0, 0) circle [x radius= 3.35, y radius= 3.35]   ;
\draw    (439.01,221.26) -- (439.01,192.13) ;
\draw [shift={(439.01,221.26)}, rotate = 270] [color={rgb, 255:red, 0; green, 0; blue, 0 }  ][fill={rgb, 255:red, 0; green, 0; blue, 0 }  ][line width=0.75]      (0, 0) circle [x radius= 3.35, y radius= 3.35]   ;
\draw    (232.68,258.12) -- (280.34,258.12) ;
\draw [shift={(282.34,258.12)}, rotate = 180] [color={rgb, 255:red, 0; green, 0; blue, 0 }  ][line width=0.75]    (10.93,-3.29) .. controls (6.95,-1.4) and (3.31,-0.3) .. (0,0) .. controls (3.31,0.3) and (6.95,1.4) .. (10.93,3.29)   ;
\draw    (232.68,250.65) -- (280.34,250.65) ;
\draw [shift={(282.34,250.65)}, rotate = 180] [color={rgb, 255:red, 0; green, 0; blue, 0 }  ][line width=0.75]    (10.93,-3.29) .. controls (6.95,-1.4) and (3.31,-0.3) .. (0,0) .. controls (3.31,0.3) and (6.95,1.4) .. (10.93,3.29)   ;
\draw    (231.66,242.43) -- (279.32,242.43) ;
\draw [shift={(281.32,242.43)}, rotate = 180] [color={rgb, 255:red, 0; green, 0; blue, 0 }  ][line width=0.75]    (10.93,-3.29) .. controls (6.95,-1.4) and (3.31,-0.3) .. (0,0) .. controls (3.31,0.3) and (6.95,1.4) .. (10.93,3.29)   ;
\draw [color={rgb, 255:red, 208; green, 2; blue, 27 }  ,draw opacity=1 ]   (306.75,261.1) -- (309.75,261.1) .. controls (311.42,259.43) and (313.08,259.43) .. (314.75,261.1) .. controls (316.42,262.77) and (318.08,262.77) .. (319.75,261.1) .. controls (321.42,259.43) and (323.08,259.43) .. (324.75,261.1) .. controls (326.42,262.77) and (328.08,262.77) .. (329.75,261.1) .. controls (331.42,259.43) and (333.08,259.43) .. (334.75,261.1) .. controls (336.42,262.77) and (338.08,262.77) .. (339.75,261.1) .. controls (341.42,259.43) and (343.08,259.43) .. (344.75,261.1) .. controls (346.42,262.77) and (348.08,262.77) .. (349.75,261.1) .. controls (351.42,259.43) and (353.08,259.43) .. (354.75,261.1) -- (355.28,261.1) -- (358.28,261.1)(306.75,264.1) -- (309.75,264.1) .. controls (311.42,262.43) and (313.08,262.43) .. (314.75,264.1) .. controls (316.42,265.77) and (318.08,265.77) .. (319.75,264.1) .. controls (321.42,262.43) and (323.08,262.43) .. (324.75,264.1) .. controls (326.42,265.77) and (328.08,265.77) .. (329.75,264.1) .. controls (331.42,262.43) and (333.08,262.43) .. (334.75,264.1) .. controls (336.42,265.77) and (338.08,265.77) .. (339.75,264.1) .. controls (341.42,262.43) and (343.08,262.43) .. (344.75,264.1) .. controls (346.42,265.77) and (348.08,265.77) .. (349.75,264.1) .. controls (351.42,262.43) and (353.08,262.43) .. (354.75,264.1) -- (355.28,264.1) -- (358.28,264.1) ;
\draw [shift={(365.28,262.6)}, rotate = 180] [color={rgb, 255:red, 208; green, 2; blue, 27 }  ,draw opacity=1 ][line width=0.75]    (10.93,-4.9) .. controls (6.95,-2.3) and (3.31,-0.67) .. (0,0) .. controls (3.31,0.67) and (6.95,2.3) .. (10.93,4.9)   ;
\draw [shift={(299.75,262.6)}, rotate = 0] [color={rgb, 255:red, 208; green, 2; blue, 27 }  ,draw opacity=1 ][line width=0.75]    (10.93,-4.9) .. controls (6.95,-2.3) and (3.31,-0.67) .. (0,0) .. controls (3.31,0.67) and (6.95,2.3) .. (10.93,4.9)   ;
\draw [color={rgb, 255:red, 208; green, 2; blue, 27 }  ,draw opacity=1 ][fill={rgb, 255:red, 255; green, 255; blue, 255 }  ,fill opacity=1 ]   (382.19,191.91) -- (382.19,194.91) .. controls (383.86,196.58) and (383.86,198.24) .. (382.19,199.91) .. controls (380.52,201.58) and (380.52,203.24) .. (382.19,204.91) .. controls (383.86,206.58) and (383.86,208.24) .. (382.19,209.91) .. controls (380.52,211.58) and (380.52,213.24) .. (382.19,214.91) .. controls (383.86,216.58) and (383.86,218.24) .. (382.19,219.91) -- (382.19,221.56) -- (382.19,224.56)(379.19,191.91) -- (379.19,194.91) .. controls (380.86,196.58) and (380.86,198.24) .. (379.19,199.91) .. controls (377.52,201.58) and (377.52,203.24) .. (379.19,204.91) .. controls (380.86,206.58) and (380.86,208.24) .. (379.19,209.91) .. controls (377.52,211.58) and (377.52,213.24) .. (379.19,214.91) .. controls (380.86,216.58) and (380.86,218.24) .. (379.19,219.91) -- (379.19,221.56) -- (379.19,224.56) ;
\draw [shift={(380.69,231.56)}, rotate = 270] [color={rgb, 255:red, 208; green, 2; blue, 27 }  ,draw opacity=1 ][line width=0.75]    (10.93,-4.9) .. controls (6.95,-2.3) and (3.31,-0.67) .. (0,0) .. controls (3.31,0.67) and (6.95,2.3) .. (10.93,4.9)   ;
\draw [shift={(380.69,184.91)}, rotate = 90] [color={rgb, 255:red, 208; green, 2; blue, 27 }  ,draw opacity=1 ][line width=0.75]    (10.93,-4.9) .. controls (6.95,-2.3) and (3.31,-0.67) .. (0,0) .. controls (3.31,0.67) and (6.95,2.3) .. (10.93,4.9)   ;
\draw    (153.83,232.74) -- (187.62,242.45) ;
\draw [shift={(187.62,242.45)}, rotate = 16.03] [color={rgb, 255:red, 0; green, 0; blue, 0 }  ][fill={rgb, 255:red, 0; green, 0; blue, 0 }  ][line width=0.75]      (0, 0) circle [x radius= 3.35, y radius= 3.35]   ;

\draw (201.96,181.46) node    {$\Tf( l_{1} ,t) \ $};
\draw (463.07,181.46) node    {$\Tf( l_{2} ,t) \ $};
\draw (340,230) node    {$\Tf(x,t) \ $};
\draw (340,190) node    {$T_{\text {amb}}( t) \ $};
\draw (380,172) node  [color={rgb, 255:red, 208; green, 2; blue, 27 }  ,opacity=1 ]  {$\mathsf{Heat\, Transfer}$};
\draw (336.41,249.69) node  [color={rgb, 255:red, 208; green, 2; blue, 27 }  ,opacity=1 ]  {$\mathsf{Diffusion} \ $};
\draw (143.22,221.58) node    {$T_{\text {in}}( t) \ $};
\draw (186.6,304.71) node    {$0$};
\draw (474.29,305.72) node    {$L$};
\draw (330.79,306.04) node    {$\Svar$};
\draw (256.28,228.51) node    {$v(t)$};

\end{tikzpicture}

%% file: Figures/Test_Bench_DG.tex
\tikzset{every picture/.style={line width=0.75pt}} 

\begin{tikzpicture}[x=0.75pt,y=0.75pt,yscale=-.7,xscale=.7]

\draw  [fill={rgb, 255:red, 126; green, 211; blue, 33 }  ,fill opacity=1 ] (392,44.61) .. controls (392,36.76) and (398.49,30.39) .. (406.5,30.39) .. controls (414.51,30.39) and (421,36.76) .. (421,44.61) .. controls (421,52.46) and (414.51,58.82) .. (406.5,58.82) .. controls (398.49,58.82) and (392,52.46) .. (392,44.61) -- cycle ;
\draw  [fill={rgb, 255:red, 155; green, 155; blue, 155 }  ,fill opacity=1 ] (499.25,44.61) .. controls (499.25,36.76) and (505.74,30.39) .. (513.75,30.39) .. controls (521.76,30.39) and (528.25,36.76) .. (528.25,44.61) .. controls (528.25,52.46) and (521.76,58.82) .. (513.75,58.82) .. controls (505.74,58.82) and (499.25,52.46) .. (499.25,44.61) -- cycle ;
\draw    (421,44.61) -- (501,44.61) ;
\draw [shift={(461,44.61)}, rotate = 180] [fill={rgb, 255:red, 0; green, 0; blue, 0 }  ][line width=0.08]  [draw opacity=0] (8.93,-4.29) -- (0,0) -- (8.93,4.29) -- cycle    ;
\draw  [fill={rgb, 255:red, 155; green, 155; blue, 155 }  ,fill opacity=1 ] (499.25,103.31) .. controls (499.25,95.46) and (505.74,89.1) .. (513.75,89.1) .. controls (521.76,89.1) and (528.25,95.46) .. (528.25,103.31) .. controls (528.25,111.16) and (521.76,117.53) .. (513.75,117.53) .. controls (505.74,117.53) and (499.25,111.16) .. (499.25,103.31) -- cycle ;
\draw  [fill={rgb, 255:red, 155; green, 155; blue, 155 }  ,fill opacity=1 ] (392,103.31) .. controls (392,95.46) and (398.49,89.1) .. (406.5,89.1) .. controls (414.51,89.1) and (421,95.46) .. (421,103.31) .. controls (421,111.16) and (414.51,117.53) .. (406.5,117.53) .. controls (398.49,117.53) and (392,111.16) .. (392,103.31) -- cycle ;
\draw  [fill={rgb, 255:red, 155; green, 155; blue, 155 }  ,fill opacity=1 ] (596,103.31) .. controls (596,95.46) and (602.49,89.1) .. (610.5,89.1) .. controls (618.51,89.1) and (625,95.46) .. (625,103.31) .. controls (625,111.16) and (618.51,117.53) .. (610.5,117.53) .. controls (602.49,117.53) and (596,111.16) .. (596,103.31) -- cycle ;
\draw  [fill={rgb, 255:red, 155; green, 155; blue, 155 }  ,fill opacity=1 ] (499.25,170.02) .. controls (499.25,162.17) and (505.74,155.8) .. (513.75,155.8) .. controls (521.76,155.8) and (528.25,162.17) .. (528.25,170.02) .. controls (528.25,177.87) and (521.76,184.24) .. (513.75,184.24) .. controls (505.74,184.24) and (499.25,177.87) .. (499.25,170.02) -- cycle ;
\draw  [fill={rgb, 255:red, 155; green, 155; blue, 155 }  ,fill opacity=1 ] (392,170.02) .. controls (392,162.17) and (398.49,155.8) .. (406.5,155.8) .. controls (414.51,155.8) and (421,162.17) .. (421,170.02) .. controls (421,177.87) and (414.51,184.24) .. (406.5,184.24) .. controls (398.49,184.24) and (392,177.87) .. (392,170.02) -- cycle ;
\draw    (528.25,44.61) -- (610.25,44.61) ;
\draw [shift={(569.25,44.61)}, rotate = 180] [fill={rgb, 255:red, 0; green, 0; blue, 0 }  ][line width=0.08]  [draw opacity=0] (8.93,-4.29) -- (0,0) -- (8.93,4.29) -- cycle    ;
\draw    (610.25,44.61) -- (610.5,87.1) ;
\draw [shift={(610.38,65.85)}, rotate = 269.65999999999997] [fill={rgb, 255:red, 0; green, 0; blue, 0 }  ][line width=0.08]  [draw opacity=0] (8.93,-4.29) -- (0,0) -- (8.93,4.29) -- cycle    ;
\draw    (513.75,58.82) -- (513.75,89.1) ;
\draw [shift={(513.75,73.96)}, rotate = 270] [fill={rgb, 255:red, 0; green, 0; blue, 0 }  ][line width=0.08]  [draw opacity=0] (8.93,-4.29) -- (0,0) -- (8.93,4.29) -- cycle    ;
\draw    (513.75,118.53) -- (513.75,155.8) ;
\draw [shift={(513.75,137.17)}, rotate = 270] [fill={rgb, 255:red, 0; green, 0; blue, 0 }  ][line width=0.08]  [draw opacity=0] (8.93,-4.29) -- (0,0) -- (8.93,4.29) -- cycle    ;
\draw    (406.5,117.53) -- (406.5,155.8) ;
\draw [shift={(406.5,136.67)}, rotate = 270] [fill={rgb, 255:red, 0; green, 0; blue, 0 }  ][line width=0.08]  [draw opacity=0] (8.93,-4.29) -- (0,0) -- (8.93,4.29) -- cycle    ;
\draw    (406.5,58.82) -- (406.5,88.1) ;
\draw [shift={(406.5,73.46)}, rotate = 270] [fill={rgb, 255:red, 0; green, 0; blue, 0 }  ][line width=0.08]  [draw opacity=0] (8.93,-4.29) -- (0,0) -- (8.93,4.29) -- cycle    ;
\draw    (610.5,118.53) -- (610.75,170.02) ;
\draw [shift={(610.63,144.27)}, rotate = 269.72] [fill={rgb, 255:red, 0; green, 0; blue, 0 }  ][line width=0.08]  [draw opacity=0] (8.93,-4.29) -- (0,0) -- (8.93,4.29) -- cycle    ;
\draw    (610.75,170.02) -- (528.25,170.02) ;
\draw [shift={(569.5,170.02)}, rotate = 360] [fill={rgb, 255:red, 0; green, 0; blue, 0 }  ][line width=0.08]  [draw opacity=0] (8.93,-4.29) -- (0,0) -- (8.93,4.29) -- cycle    ;
\draw    (499.25,170.02) -- (421,170.02) ;
\draw [shift={(460.13,170.02)}, rotate = 360] [fill={rgb, 255:red, 0; green, 0; blue, 0 }  ][line width=0.08]  [draw opacity=0] (8.93,-4.29) -- (0,0) -- (8.93,4.29) -- cycle    ;
\draw    (364,16.67) -- (392.33,35.18) ;
\draw [shift={(394,36.27)}, rotate = 213.17000000000002] [color={rgb, 255:red, 0; green, 0; blue, 0 }  ][line width=0.75]    (10.93,-3.29) .. controls (6.95,-1.4) and (3.31,-0.3) .. (0,0) .. controls (3.31,0.3) and (6.95,1.4) .. (10.93,3.29)   ;
\draw    (528.25,103.31) -- (556.58,121.83) ;
\draw [shift={(558.25,122.92)}, rotate = 213.17000000000002] [color={rgb, 255:red, 0; green, 0; blue, 0 }  ][line width=0.75]    (10.93,-3.29) .. controls (6.95,-1.4) and (3.31,-0.3) .. (0,0) .. controls (3.31,0.3) and (6.95,1.4) .. (10.93,3.29)   ;
\draw    (625,103.31) -- (653.33,121.83) ;
\draw [shift={(655,122.92)}, rotate = 213.17000000000002] [color={rgb, 255:red, 0; green, 0; blue, 0 }  ][line width=0.75]    (10.93,-3.29) .. controls (6.95,-1.4) and (3.31,-0.3) .. (0,0) .. controls (3.31,0.3) and (6.95,1.4) .. (10.93,3.29)   ;
\draw    (421,103.31) -- (449.33,121.83) ;
\draw [shift={(451,122.92)}, rotate = 213.17000000000002] [color={rgb, 255:red, 0; green, 0; blue, 0 }  ][line width=0.75]    (10.93,-3.29) .. controls (6.95,-1.4) and (3.31,-0.3) .. (0,0) .. controls (3.31,0.3) and (6.95,1.4) .. (10.93,3.29)   ;
\draw    (365.67,144.15) -- (394,162.67) ;
\draw [shift={(364,143.06)}, rotate = 33.17] [color={rgb, 255:red, 0; green, 0; blue, 0 }  ][line width=0.75]    (10.93,-3.29) .. controls (6.95,-1.4) and (3.31,-0.3) .. (0,0) .. controls (3.31,0.3) and (6.95,1.4) .. (10.93,3.29)   ;
\draw    (120,40) -- (279,40) ;
\draw    (120,172) -- (279,172) ;
\draw    (120,116) -- (120,172) ;
\draw [shift={(120,172)}, rotate = 90] [color={rgb, 255:red, 0; green, 0; blue, 0 }  ][fill={rgb, 255:red, 0; green, 0; blue, 0 }  ][line width=0.75]      (0, 0) circle [x radius= 3.35, y radius= 3.35]   ;
\draw   (119.63,101) -- (126.5,116) -- (112.75,116) -- cycle ;
\draw   (119.62,101) -- (112.76,86) -- (126.51,86.01) -- cycle ;
\draw    (110,106.75) -- (132.23,94.94) ;
\draw [shift={(134,94)}, rotate = 512.02] [color={rgb, 255:red, 0; green, 0; blue, 0 }  ][line width=0.75]    (10.93,-3.29) .. controls (6.95,-1.4) and (3.31,-0.3) .. (0,0) .. controls (3.31,0.3) and (6.95,1.4) .. (10.93,3.29)   ;

\draw    (120,39) -- (120,87) ;
\draw [shift={(120,39)}, rotate = 90] [color={rgb, 255:red, 0; green, 0; blue, 0 }  ][fill={rgb, 255:red, 0; green, 0; blue, 0 }  ][line width=0.75]      (0, 0) circle [x radius= 3.35, y radius= 3.35]   ;
\draw    (135.15,75.13) -- (120,75.18) ;
\draw [shift={(120,75.18)}, rotate = 179.81] [color={rgb, 255:red, 0; green, 0; blue, 0 }  ][fill={rgb, 255:red, 0; green, 0; blue, 0 }  ][line width=0.75]      (0, 0) circle [x radius= 3.35, y radius= 3.35]   ;
\draw   (146.12,74.85) -- (135.13,80.89) -- (135.17,68.7) -- cycle ;
\draw   (146.12,74.85) -- (157.12,68.81) -- (157.07,81) -- cycle ;
\draw    (146.12,74.85) -- (146.12,67.2) ;
\draw   (142.59,66.79) .. controls (142.47,66.38) and (142.41,65.95) .. (142.4,65.5) .. controls (142.37,63.09) and (144.01,61.11) .. (146.06,61.07) .. controls (148.11,61.03) and (149.8,62.95) .. (149.83,65.36) .. controls (149.84,65.81) and (149.79,66.25) .. (149.68,66.66) -- cycle ;
\draw    (168,75.14) -- (157.09,75.18) ;
\draw [shift={(171,75.13)}, rotate = 179.79] [fill={rgb, 255:red, 0; green, 0; blue, 0 }  ][line width=0.08]  [draw opacity=0] (8.93,-4.29) -- (0,0) -- (8.93,4.29) -- cycle    ;
\draw    (202,115) -- (202,172) ;
\draw [shift={(202,172)}, rotate = 90] [color={rgb, 255:red, 0; green, 0; blue, 0 }  ][fill={rgb, 255:red, 0; green, 0; blue, 0 }  ][line width=0.75]      (0, 0) circle [x radius= 3.35, y radius= 3.35]   ;
\draw   (201.63,101) -- (208.5,116) -- (194.75,116) -- cycle ;
\draw   (201.62,101) -- (194.76,86) -- (208.51,86.01) -- cycle ;
\draw    (192,106.75) -- (214.23,94.94) ;
\draw [shift={(216,94)}, rotate = 512.02] [color={rgb, 255:red, 0; green, 0; blue, 0 }  ][line width=0.75]    (10.93,-3.29) .. controls (6.95,-1.4) and (3.31,-0.3) .. (0,0) .. controls (3.31,0.3) and (6.95,1.4) .. (10.93,3.29)   ;

\draw    (202,39) -- (202,86) ;
\draw [shift={(202,39)}, rotate = 90] [color={rgb, 255:red, 0; green, 0; blue, 0 }  ][fill={rgb, 255:red, 0; green, 0; blue, 0 }  ][line width=0.75]      (0, 0) circle [x radius= 3.35, y radius= 3.35]   ;
\draw    (217.15,75.13) -- (202,75.18) ;
\draw [shift={(202,75.18)}, rotate = 179.81] [color={rgb, 255:red, 0; green, 0; blue, 0 }  ][fill={rgb, 255:red, 0; green, 0; blue, 0 }  ][line width=0.75]      (0, 0) circle [x radius= 3.35, y radius= 3.35]   ;
\draw   (228.12,74.85) -- (217.13,80.89) -- (217.17,68.7) -- cycle ;
\draw   (228.12,74.85) -- (239.12,68.81) -- (239.07,81) -- cycle ;
\draw    (228.12,74.85) -- (228.12,67.2) ;
\draw   (224.59,66.79) .. controls (224.47,66.38) and (224.41,65.95) .. (224.4,65.5) .. controls (224.37,63.09) and (226.01,61.11) .. (228.06,61.07) .. controls (230.11,61.03) and (231.8,62.95) .. (231.83,65.36) .. controls (231.84,65.81) and (231.79,66.25) .. (231.68,66.66) -- cycle ;
\draw    (250,75.14) -- (239.09,75.18) ;
\draw [shift={(253,75.13)}, rotate = 179.79] [fill={rgb, 255:red, 0; green, 0; blue, 0 }  ][line width=0.08]  [draw opacity=0] (8.93,-4.29) -- (0,0) -- (8.93,4.29) -- cycle    ;
\draw    (279,116) -- (279,172) ;
\draw   (278.63,101) -- (285.5,116) -- (271.75,116) -- cycle ;
\draw   (278.62,101) -- (271.76,86) -- (285.51,86.01) -- cycle ;
\draw    (269,106.75) -- (291.23,94.94) ;
\draw [shift={(293,94)}, rotate = 512.02] [color={rgb, 255:red, 0; green, 0; blue, 0 }  ][line width=0.75]    (10.93,-3.29) .. controls (6.95,-1.4) and (3.31,-0.3) .. (0,0) .. controls (3.31,0.3) and (6.95,1.4) .. (10.93,3.29)   ;

\draw    (279,39) -- (279,87) ;
\draw    (294.15,75.13) -- (279,75.18) ;
\draw [shift={(279,75.18)}, rotate = 179.81] [color={rgb, 255:red, 0; green, 0; blue, 0 }  ][fill={rgb, 255:red, 0; green, 0; blue, 0 }  ][line width=0.75]      (0, 0) circle [x radius= 3.35, y radius= 3.35]   ;
\draw   (305.12,74.85) -- (294.13,80.89) -- (294.17,68.7) -- cycle ;
\draw   (305.12,74.85) -- (316.12,68.81) -- (316.07,81) -- cycle ;
\draw    (305.12,74.85) -- (305.12,67.2) ;
\draw   (301.59,66.79) .. controls (301.47,66.38) and (301.41,65.95) .. (301.4,65.5) .. controls (301.37,63.09) and (303.01,61.11) .. (305.06,61.07) .. controls (307.11,61.03) and (308.8,62.95) .. (308.83,65.36) .. controls (308.84,65.81) and (308.79,66.25) .. (308.68,66.66) -- cycle ;

\draw    (327,75.14) -- (316.09,75.18) ;
\draw [shift={(330,75.13)}, rotate = 179.79] [fill={rgb, 255:red, 0; green, 0; blue, 0 }  ][line width=0.08]  [draw opacity=0] (8.93,-4.29) -- (0,0) -- (8.93,4.29) -- cycle    ;
\draw    (47,40) -- (119,40) ;
\draw    (47,40) -- (32.5,54.5) ;
\draw    (47,40) -- (32.5,25.5) ;
\draw   (76.5,22.75) -- (91.5,40) -- (76.5,57.25) -- (61.5,40) -- cycle ;
\draw    (76.5,22.75) -- (76.5,38) ;
\draw [shift={(76.5,40)}, rotate = 270] [color={rgb, 255:red, 0; green, 0; blue, 0 }  ][line width=0.75]    (10.93,-3.29) .. controls (6.95,-1.4) and (3.31,-0.3) .. (0,0) .. controls (3.31,0.3) and (6.95,1.4) .. (10.93,3.29)   ;
\draw    (76.5,57.25) -- (76.5,42) ;
\draw [shift={(76.5,40)}, rotate = 450] [color={rgb, 255:red, 0; green, 0; blue, 0 }  ][line width=0.75]    (10.93,-3.29) .. controls (6.95,-1.4) and (3.31,-0.3) .. (0,0) .. controls (3.31,0.3) and (6.95,1.4) .. (10.93,3.29)   ;
\draw    (11,172) -- (120,172) ;
\draw    (11,40) -- (18,40) ;
\draw    (120,134.97) -- (129,134.86) ;
\draw    (134.02,134.75) -- (147.04,134.63) ;
\draw [shift={(134.02,134.75)}, rotate = 359.48] [color={rgb, 255:red, 0; green, 0; blue, 0 }  ][fill={rgb, 255:red, 0; green, 0; blue, 0 }  ][line width=0.75]      (0, 0) circle [x radius= 2.68, y radius= 2.68]   ;

\draw    (120.2,40) -- (120.11,31) ;
\draw    (120.01,25.98) -- (119.92,12.96) ;
\draw [shift={(120.01,25.98)}, rotate = 269.62] [color={rgb, 255:red, 0; green, 0; blue, 0 }  ][fill={rgb, 255:red, 0; green, 0; blue, 0 }  ][line width=0.75]      (0, 0) circle [x radius= 2.68, y radius= 2.68]   ;
\draw    (202,134.97) -- (211,134.86) ;

\draw    (216.02,134.75) -- (229.04,134.63) ;
\draw [shift={(216.02,134.75)}, rotate = 359.48] [color={rgb, 255:red, 0; green, 0; blue, 0 }  ][fill={rgb, 255:red, 0; green, 0; blue, 0 }  ][line width=0.75]      (0, 0) circle [x radius= 2.68, y radius= 2.68]   ;

\draw    (279,134.97) -- (288,134.86) ;

\draw    (293.02,134.75) -- (306.04,134.63) ;
\draw [shift={(293.02,134.75)}, rotate = 359.48] [color={rgb, 255:red, 0; green, 0; blue, 0 }  ][fill={rgb, 255:red, 0; green, 0; blue, 0 }  ][line width=0.75]      (0, 0) circle [x radius= 2.68, y radius= 2.68]   ;
\draw    (100.2,172) -- (100.11,163) ;
\draw    (100.01,157.98) -- (99.92,144.96) ;
\draw [shift={(100.01,157.98)}, rotate = 269.62] [color={rgb, 255:red, 0; green, 0; blue, 0 }  ][fill={rgb, 255:red, 0; green, 0; blue, 0 }  ][line width=0.75]      (0, 0) circle [x radius= 2.68, y radius= 2.68]   ;

\draw   (18,40) .. controls (18,31.99) and (24.49,25.5) .. (32.5,25.5) .. controls (40.51,25.5) and (47,31.99) .. (47,40) .. controls (47,48.01) and (40.51,54.5) .. (32.5,54.5) .. controls (24.49,54.5) and (18,48.01) .. (18,40) -- cycle ;

\draw   (90,153.07) .. controls (89.95,147.54) and (94.39,143.01) .. (99.93,142.96) .. controls (105.46,142.91) and (109.99,147.35) .. (110.04,152.89) .. controls (110.09,158.42) and (105.65,162.95) .. (100.11,163) .. controls (94.58,163.05) and (90.05,158.61) .. (90,153.07) -- cycle ;
\draw   (110,21.07) .. controls (109.95,15.54) and (114.39,11.01) .. (119.93,10.96) .. controls (125.46,10.91) and (129.99,15.35) .. (130.04,20.89) .. controls (130.09,26.42) and (125.65,30.95) .. (120.11,31) .. controls (114.58,31.05) and (110.05,26.61) .. (110,21.07) -- cycle ;
\draw   (138.9,124.73) .. controls (144.43,124.66) and (148.97,129.09) .. (149.04,134.63) .. controls (149.1,140.16) and (144.67,144.7) .. (139.14,144.76) .. controls (133.6,144.83) and (129.07,140.4) .. (129,134.86) .. controls (128.93,129.33) and (133.37,124.79) .. (138.9,124.73) -- cycle ;
\draw   (220.9,124.73) .. controls (226.43,124.66) and (230.97,129.09) .. (231.04,134.63) .. controls (231.1,140.16) and (226.67,144.7) .. (221.14,144.76) .. controls (215.6,144.83) and (211.06,140.4) .. (211,134.86) .. controls (210.93,129.33) and (215.37,124.79) .. (220.9,124.73) -- cycle ;
\draw   (297.9,124.73) .. controls (303.43,124.66) and (307.97,129.09) .. (308.04,134.63) .. controls (308.1,140.16) and (303.67,144.7) .. (298.14,144.76) .. controls (292.6,144.83) and (288.07,140.4) .. (288,134.86) .. controls (287.93,129.33) and (292.37,124.79) .. (297.9,124.73) -- cycle ;

\draw (400-3,37.52-1) node [anchor=north west][inner sep=0.75pt]  [font=\scriptsize]  {$h_{1}$};
\draw (504.25,37.52-1) node [anchor=north west][inner sep=0.75pt]  [font=\scriptsize]  {$h_{5}$};
\draw (400-3,97.22-1) node [anchor=north west][inner sep=0.75pt]  [font=\scriptsize]  {$h_{2}$};
\draw (507.25-3,97.22-1) node [anchor=north west][inner sep=0.75pt]  [font=\scriptsize]  {$h_{3}$};
\draw (604-3,97.22-1) node [anchor=north west][inner sep=0.75pt]  [font=\scriptsize]  {$h_{4}$};
\draw (400-3,163.93-1) node [anchor=north west][inner sep=0.75pt]  [font=\scriptsize]  {$h_{7}$};
\draw (507.25-3,163.93-1) node [anchor=north west][inner sep=0.75pt]  [font=\scriptsize]  {$h_{6}$};
\draw (449.5,22.93-1) node [anchor=north west][inner sep=0.75pt]  [font=\scriptsize]  {$\textcolor[rgb]{0.82,0.01,0.11}{q}\textcolor[rgb]{0.82,0.01,0.11}{_{\textcolor[rgb]{0.82,0.01,0.11}{7}}}$};
\draw (414,63.58) node [anchor=north west][inner sep=0.75pt]  [font=\scriptsize]  {$\textcolor[rgb]{0.82,0.01,0.11}{q}\textcolor[rgb]{0.82,0.01,0.11}{_{1}}$};
\draw (414,128.37) node [anchor=north west][inner sep=0.75pt]  [font=\scriptsize]  {$\textcolor[rgb]{0.82,0.01,0.11}{q}\textcolor[rgb]{0.82,0.01,0.11}{_{4}}$};
\draw (524,128.37) node [anchor=north west][inner sep=0.75pt]  [font=\scriptsize]  {$\textcolor[rgb]{0.82,0.01,0.11}{q}\textcolor[rgb]{0.82,0.01,0.11}{_{5}}$};
\draw (618,128.37) node [anchor=north west][inner sep=0.75pt]  [font=\scriptsize]  {$\textcolor[rgb]{0.82,0.01,0.11}{q}\textcolor[rgb]{0.82,0.01,0.11}{_{6}}$};
\draw (523,63.58) node [anchor=north west][inner sep=0.75pt]  [font=\scriptsize]  {$\textcolor[rgb]{0.82,0.01,0.11}{q}\textcolor[rgb]{0.82,0.01,0.11}{_{2}}$};
\draw (452.5,149.62) node [anchor=north west][inner sep=0.75pt]  [font=\scriptsize]  {$\textcolor[rgb]{0.82,0.01,0.11}{q}\textcolor[rgb]{0.82,0.01,0.11}{_{8}}$};
\draw (616,63.58) node [anchor=north west][inner sep=0.75pt]  [font=\scriptsize]  {$\textcolor[rgb]{0.82,0.01,0.11}{q}\textcolor[rgb]{0.82,0.01,0.11}{_{3}}$};
\draw (641+2,95.8) node [anchor=north west][inner sep=0.75pt]  [font=\scriptsize]  {$\textcolor[rgb]{0.29,0.56,0.89}{q}\textcolor[rgb]{0.29,0.56,0.89}{_{3}^{d}}$};
\draw (541+2,95.8) node [anchor=north west][inner sep=0.75pt]  [font=\scriptsize]  {$\textcolor[rgb]{0.29,0.56,0.89}{q}\textcolor[rgb]{0.29,0.56,0.89}{_{2}^{d}}$};
\draw (432+2,95.8) node [anchor=north west][inner sep=0.75pt]  [font=\scriptsize]  {$\textcolor[rgb]{0.29,0.56,0.89}{q}\textcolor[rgb]{0.29,0.56,0.89}{_{1}^{d}}$};
\draw (365,5.4) node [anchor=north west][inner sep=0.75pt]  [font=\scriptsize]  {$\textcolor[rgb]{0.49,0.83,0.13}{q}\textcolor[rgb]{0.49,0.83,0.13}{_{\mathrm{up}}}$};
\draw (365,128.65) node [anchor=north west][inner sep=0.75pt]  [font=\scriptsize]  {$\textcolor[rgb]{0.29,0.56,0.89}{q}\textcolor[rgb]{0.29,0.56,0.89}{_{\mathrm{out}}}$};
\draw (13,62) node [anchor=north west][inner sep=0.75pt]  [font=\scriptsize] [align=left] {pump};
\draw (61,8) node [anchor=north west][inner sep=0.75pt]  [font=\scriptsize] [align=left] {heater};
\draw (137,10.4) node [anchor=north west][inner sep=0.75pt]  [font=\scriptsize]  {$T_{\mathrm{up}} ,h_{\mathrm{up}} \ \ $};
\draw (150,118.4) node [anchor=north west][inner sep=0.75pt]  [font=\scriptsize]  {$T_{1} \ $};
\draw (232,117.4) node [anchor=north west][inner sep=0.75pt]  [font=\scriptsize]  {$T_{2} \ $};
\draw (310,118.4) node [anchor=north west][inner sep=0.75pt]  [font=\scriptsize]  {$T_{3} \ $};
\draw (44,142.4) node [anchor=north west][inner sep=0.75pt]  [font=\scriptsize]  {$T_{\mathrm{d}} ,h_{\mathrm{d}} \ \ $};

\end{tikzpicture}

%% file: Paper_Main_IEEE_CP_Notice.bbl
\begin{thebibliography}{10}
\providecommand{\url}[1]{#1}
\csname url@rmstyle\endcsname
\providecommand{\newblock}{\relax}
\providecommand{\bibinfo}[2]{#2}
\providecommand\BIBentrySTDinterwordspacing{\spaceskip=0pt\relax}
\providecommand\BIBentryALTinterwordstretchfactor{4}
\providecommand\BIBentryALTinterwordspacing{\spaceskip=\fontdimen2\font plus
\BIBentryALTinterwordstretchfactor\fontdimen3\font minus
  \fontdimen4\font\relax}
\providecommand\BIBforeignlanguage[2]{{%
\expandafter\ifx\csname l@#1\endcsname\relax
\typeout{** WARNING: IEEEtran.bst: No hyphenation pattern has been}%
\typeout{** loaded for the language `#1'. Using the pattern for}%
\typeout{** the default language instead.}%
\else
\language=\csname l@#1\endcsname
\fi
#2}}

\bibitem{Review_Tools}
B.~Coelho and A.~Andrade-Campos, ``Efficiency achievement in water supply
  systems—a review,'' \emph{Renewable and Sustainable Energy Reviews},
  vol.~30, pp. 59--84, 2014.

\bibitem{rossman1999epanet}
L.~A. Rossman, ``The epanet programmer's toolkit for analysis of water
  distribution systems,'' in \emph{WRPMD'99: Preparing for the 21st Century},
  1999, pp. 1--10.

\bibitem{WQ_Epanet}
A.~Seyoum and T.~Tanyimboh, ``Integration of hydraulic and water quality
  modelling in distribution networks: Epanet-pmx,'' \emph{Water Resources
  Management}, vol.~31, 11 2017.

\bibitem{RTP_Epanet}
S.~Hatchett, J.~Uber, D.~Boccelli, T.~Haxton, R.~Janke, A.~Kramer, A.~Matracia,
  and S.~Panguluri, ``Real-time distribution system modeling: development,
  application and insights,'' \emph{Computing and Control for the Water
  Industry 2011 - CCWI 2011Computing and Control for the Water Industry 2011 -
  CCWI 2011}, 05 2011.

\bibitem{RTP_Epanet_1}
P.~Ingeduld, ``Real time forecasting with epanet,'' 05 2007.

\bibitem{RTC_Review}
E.~Creaco, A.~Campisano, N.~Fontana, G.~Marini, P.~Page, and T.~Walski, ``Real
  time control of water distribution networks: A state-of-the-art review,''
  \emph{Water Research}, vol. 161, pp. 517--530, 2019.

\bibitem{WQModels4MPC}
S.~Wang, A.~F. Taha, and A.~A. Abokifa, ``How effective is model predictive
  control in real-time water quality regulation? state-space modeling and
  scalable control,'' \emph{Water Resources Research}, vol.~57, no.~5, 2021.

\bibitem{WQMOR}
S.~Wang, A.~F. Taha, A.~Chakrabarty, L.~Sela, and A.~Abokifa, ``Model order
  reduction for water quality dynamics,'' 2021.

\bibitem{MORParamOverview}
P.~Benner, S.~Gugercin, and K.~Willcox, ``A survey of projection-based model
  reduction methods for parametric dynamical systems,'' \emph{SIAM Review},
  vol.~57, no.~4, pp. 483--531, 2015.

\bibitem{POD_MOR}
S.~Volkwein, ``Proper orthogonal decomposition: Theory and reduced-order
  modelling,'' \emph{Lecture Notes, University of Konstanz}, 01 2012.

\bibitem{TPWL_M}
B.~N. Bond and L.~Daniel, ``A piecewise-linear moment-matching approach to
  parameterized model-order reduction for highly nonlinear systems,''
  \emph{IEEE Transactions on Computer-Aided Design of Integrated Circuits and
  Systems}, vol.~26, no.~12, pp. 2116--2129, 2007.

\bibitem{MORLin}
H.~Panzer, \emph{Model order reduction by Krylov subspace methods with global
  error bounds and automatic choice of parameters: Zugl.: M{\"u}nchen, Techn.
  Univ., Diss., 2014}, ser. Ingenieurwissenschaften.\hskip 1em plus 0.5em minus
  0.4em\relax M{\"u}nchen: {Dr. Hut}, 2014.

\bibitem{MOROverview1}
{Ulrike Baur}, {Peter Benner}, and {Lihong Feng}, ``Model order reduction for
  linear and nonlinear systems: A system-theoretic perspective,''
  \emph{Archives of Computational Methods in Engineering}, vol.~21, no.~4, pp.
  331--358, 2014.

\bibitem{MOROverview2}
C.~Beattie and S.~Gugercin, ``Model reduction by rational interpolation,''
  2014.

\bibitem{MORbilinH2Zhang.2002}
L.~Zhang and J.~Lam, ``On h2 model reduction of bilinear systems,''
  \emph{Automatica}, vol.~38, no.~2, pp. 205--216, 2002.

\bibitem{MORQb}
P.~Benner, P.~Goyal, and S.~Gugercin, ``$\mathcal{H}_2$-quasi-optimal model
  order reduction for quadratic-bilinear control systems.''

\bibitem{MORpoly}
P.~Benner and P.~Goyal, ``Interpolation-based model order reduction for
  polynomial parametric systems.''

\bibitem{Elkhashap.2021b}
A.~Elkhashap, D.~Rüschen, and D.~Abel, ``Distributed parameter modeling of
  fluid transmission lines,'' \emph{Journal of Process Control}, 2021, in
  press.

\bibitem{Elkhashap2022model}
A.~Elkhashap and D.~Abel, ``Model order reduction of
  advection-dispersion-reaction equation with time-varying coefficients,
  application to real-time water quality monitoring,'' in \emph{2022 European
  Control Conference (ECC)}, 2022.

\bibitem{Hamdi.2007}
S.~Hamdi, W.~Schiesser, and G.~Griffiths, ``Method of lines,''
  \emph{Scholarpedia}, vol.~2, no.~7, p. 2859, 2007.

\bibitem{Abgrall.2017}
R.~Abgrall, C.-W. Shu, and Q.~Du, Eds., \emph{Handbook of numerical methods for
  hyperbolic problems: Applied and modern issues}, ser. Handbook of numerical
  analysis.\hskip 1em plus 0.5em minus 0.4em\relax Amsterdam, The Netherlands
  and Kidlington, Oxford, United Kingdom: {North-Holland an imprint of
  Elsevier}, 2017, vol. volume 18.

\bibitem{TensorDec}
S.~Rabanser, O.~Shchur, and S.~G{\"u}nnemann, ``Introduction to tensor
  decompositions and their applications in machine learning,'' 2017.

\bibitem{MORbilinH2}
P.~Benner and T.~Breiten, ``Interpolation-based $\mathcal{H}_2$-model reduction
  of bilinear control systems,'' \emph{SIAM Journal on Matrix Analysis and
  Applications}, vol.~33, no.~3, pp. 859--885, 2012.

\bibitem{Elkhashap.2021a}
A.~Elkhashap and D.~Abel, ``Parametric model order reduction of variable
  parameter axial dispersion model,'' \emph{Conference on Control Technology
  and Applications (CCTA)}, 2021.

\bibitem{Andersson2019}
J.~A.~E. Andersson, J.~Gillis, G.~Horn, J.~B. Rawlings, and M.~Diehl,
  ``{CasADi} -- {A} software framework for nonlinear optimization and optimal
  control,'' \emph{Mathematical Programming Computation}, vol.~11, no.~1, pp.
  1--36, 2019.

\bibitem{pimenta2018performance}
B.~D. Pimenta, A.~D. Robaina, M.~X. Peiter, W.~Mezzomo, J.~H. Kirchner, and
  L.~H. Ben, ``Performance of explicit approximations of the coefficient of
  head loss for pressurized conduits,'' \emph{Revista Brasileira de Engenharia
  Agr{\'\i}cola e Ambiental}, vol.~22, no.~5, pp. 301--307, 2018.

\bibitem{todini1988gradient}
E.~Todini and S.~Pilati, ``A gradient algorithm for the analysis of pipe
  networks,'' in \emph{Computer applications in water supply: vol. 1---systems
  analysis and simulation}.\hskip 1em plus 0.5em minus 0.4em\relax Research
  Studies Press Ltd., 1988, pp. 1--20.

\bibitem{GGA2011}
O.~Giustolisi, L.~Berardi, and D.~Laucelli, ``{Generalizing WDN simulation
  models to variable tank levels},'' \emph{Journal of Hydroinformatics},
  vol.~14, no.~3, pp. 562--573, 11 2011.

\bibitem{MATLAB_2020}
MATLAB, \emph{version 9.9.0 (R2020b)}.\hskip 1em plus 0.5em minus 0.4em\relax
  Natick, Massachusetts: The MathWorks Inc., 2020.

\end{thebibliography}
